# A Low Overhead Cooperative-based Authentication Protocol for VANETs

Vahid Ranjbar[1], Ali Mohammad Afshin Hemmatyar[2]

[1]University of Tehran, [2]Sharif University of Technology,
Emails: vranjbar@ut.ac.ir, hemmatyar@sharif.edu

*Abstract*—Vehicular ad-hoc networks (VANETs) have been proposed to automate transportation industry in order to increase its accuracy, efficiency, throughput, and specially safety. Security plays an Undeniable important role on implementing VANETs in real life. Authentication is one of the basic elements of VANETs security. Proposed authentications protocols suffer from high overhead and cost. This paper presents a computation division based authentication which divide signature approvals between neighbor vehicles consequently decrease vehicles computation load. Simulation shows presented protocol propose an almost constant latency and closely zero message loss ratio related to traffic load, and improved efficiency compared with GSIS protocol.

## I. INTRODUCTION

Nowadays, the transportation industry has become one of the most important subjects of social affairs. Vehicular ad-hoc networks (VANETs) have been proposed to automate this industry and decrease its risk for human lives. These networks can be used to increase the efficiency of driver assistance advanced systems, safety and capacity of roads, and comfort level of driver and passengers. VANET applications can be divided in welfare and safety groups. Traffic management information, electronic payment systems, navigation improvement, and providing welfare information and entertainment for passengers can be considered as welfare ones, and Accident preventing and collaboration with relief and security vehicles are the safety group.

In general, vehicular ad-hoc networks can be considered as mobile ad-hoc networks. In a VANET, each vehicle operates as a smart node and can connect to other vehicles and transportation equipment. VANETs may have infrastructure but there is usually no infrastructure in mobile ad-hoc networks. Also, VANETs have more dynamic topology and higher speed of nodes. Path limitation for vehicles, existence of tools to predict their motion, nonexistence of limitation for transmitted power and consumed energy by the network, and considerable nodes density variation related to area are the VANETs difference.

One of the most important goal of VANETs is to increase the safety of transportation industry. To achieve this goal,

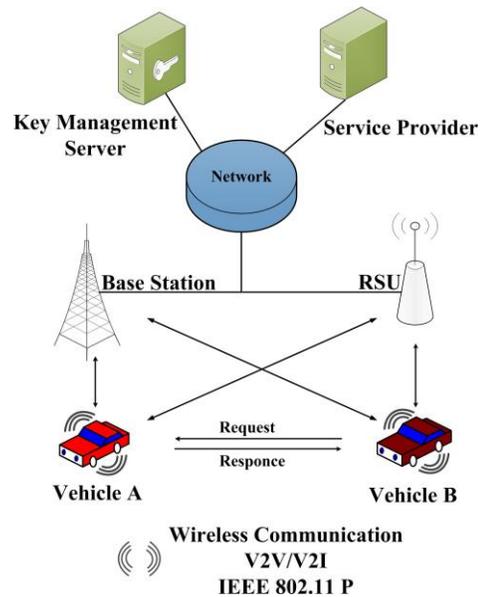

Fig. 1. Schematic for an ID-based method.

VANETs themselves must be accurate, secure, and safe in the first step. Any unintentional or intentional mistakes could be followed by irreversible results.

Security protocols that are proposed to obtain security in VANETs must satisfy requirements including authentication, privacy protection and anonymity, ability to track malicious nodes and revoking their certificate, preventing denial of service, confidentiality, integrity, and efficiency. Obviously, there should be a tradeoff between these security requirements and the cost and complexity of protocol. Authentication is one of the fundamental elements of VANETs security that verify messages transmitting and receiving by network's legitimate authorized nodes. Authentication also decrease illegal nodes attacks [1]. Privacy protection and anonymity prevent vehicles path tracking and privacy abuses. So, the proposed authentication model should provide appropriate level of anonymity [1]. Impossibility of linking data prevent tracking a vehicle by enemy through eavesdropping network messages [2].

the possibility of identification and removal of malicious vehicle. Impossibility of denial of service prevent transmitter or receiver form denying that transmitting or receiving [3], [4]. So, false alarms are prevented and it will be possible to prove that a false alarm caused an accident [5]. Confidentiality and integrity grantee that other members specially a malicious one can not change a message [6]. The efficiency considers the protocol costs and latency requirements [2].

As VANETs have no infrastructure and subsequently a center node for network communication management and routing, nodes should perform routing, communication management and network topology configuration themselves. This can helps an attacker node to attack and destroy the network. Despite of low possibility of public key infrastructure and building VANETs for a short time premeditated plane, they need a key distribution and authentication protocol to prevent an attacker node entering.

Security of vehicular ad-hoc networks is one of the major problems that prevent implementation of VANETs in recent years. Many solutions such as information encryption, network isolation, using authentication for new entering vehicles, applying hardware with more than 99.99 percent reliability are proposed to improve VANETs security. But, still there are many problems should be solved to make VANETs implementable.

One of the most important elements of VANETs security is authentication of transmitted messages by vehicles to prevent unauthorized nodes entering and external attacks. Popular authentication protocols can be divided in two groups. Some protocols use public key infrastructure for authentication [7]–[9], and the others include Id-based encryption [10]–[12]. At the first group each vehicle receives a pair of public and private key and a valid certificate from a reliable center. Main drawback of this protocols is high calculation overhead due to public key that cause messages latency. In the second group of authentication that use id-based encryption, each vehicle has an exclusive Id. Key management center build a private key corresponding vehicle's Id and transmit it to vehicle over a secure channel. Lower calculation overhead and easier signature approval are two main advantages of this protocols towards public key protocols. On the other hand, single point of failure (key management center) and possibility of Id forgery are disadvantages of the second group [13]–[16]. All above protocols are proposed without considering privacy and certificate revocation requirements. To satisfying certificate revocation requirement "certificate revoking list" is used that include all violator vehicles. This list is saved in all vehicles and used to prevent receiving a message from a violator vehicle. Two major problem of this solution is cost of keeping the list up-dated and saving it in all vehicles [17], [18].

Some protocols presented to solve the privacy protection and anonymity problems use "pseudonym" for user anonymity [19], [20]. Regarding this, each user receives several temporary certificates from a reliable center that are stored with user's real profile in a reliable database. This protocols use almost more than 43 thousands temporary certificates over just one year to propose an acceptable level of efficiency and anonymity [17]. This amount of certificates is followed by problems such as memory overhead, long searching time, long certificate revoking time, up-dating cost, and high communication overhead.

TABLE I
ADVANTAGES AND DISADVANTAGES OF RELATED PROTOCOLS.

| Protocols | Advantages | Disadvantages |
|---|---|---|
| GSIS [11] | User authentication and anonymity, Tracking ability | High loss rate, Approval time increasment proportional to revoked certificates |
| Khomejani [22] | User authentication and anonymity, No need for road side unit | High loss rate, Tracking disability, Transmitter can deny message transmitting |
| GAP [23] | User authentication and anonymity, Tracking ability | High loss rate |

Group signature could be used for users' anonymity in order to remove temporary certificate weaknesses [9], [21]. In this protocol, group manager authenticates each user with its real identity at the first step. Also, the manager produces a private key for user and gives it to user with group public key After Authentication and saves the user private key and real identity in its database. In this protocol, each group manager has a master key which can be used with user signature to obtain user private key, then it can search its database to find message transmitter real identity. Hybrid protocols combine this two protocols to use their advantages together [11]. Advantages and drawbacks of three related protocols are summarized in Table I.

## II. PROPOSED PROTOCOL

In the proposed protocol, it is assumed that there is a reliable center for key distribution and management. Also, it assumes that roadside units (RSUs) have a high computing abilities and can build a proper communication coverage to up-date and revoke the certificates. It is assumed roadside units use an electronic elliptical curve signature algorithm as public key infrastructure for certification. This algorithm is usually used in other VANETs authentication protocols. In proposed protocol, each RSU is considered as a local reliable center on its zone. So, vehicles can receive local valid certificate for each zone from its RSU. Country transportation centers can act as backbone of this protocol.

The protocol assumed each vehicle can at-least connect to one roadside unit to up-date its temporary certificate. Also, RSUs are physically protected and secure, connected to other units and reliable center over a secure Internet, satellite, radio waves or cable base channel. Each vehicles is loaded with an accurate navigation system. Moreover, it's assumed that attacker can access and change all messages, so confidentiality

TABLE II
ABBREVIATIONS

| Abbreviations | Explanation |
| --- | --- |
| $Encrypt_K(M)$ | Message M asymmetric encryption algorithm by key K |
| $Decrypt_K(M)$ | Message M asymmetric decryption algorithm by key K |
| $sign_{K^{-1}}(M)$ | Simple electronically signing algorithm for message M using private key $K^{-1}$ |
| $Cert_i$ | Vehicle i certificate signed by reliable center |
| $P\,Key_S^{-1}$ | Roadside unit/ Vehicle (S) private key for electronically signing |
| $P\,Key_S^{+1}$ | Roadside unit/ Vehicle S public key for electronically signing |
| $RL$ | Revoked certificates list (Revoking List) |
| $K_S^{-1}$ | Vehicle S private key for temporary certificate |
| $K_S^{+1}$ | Vehicle S private key for temporary certificate |

1) OBU: $(K_S^{+1}, K_S^{-1}) \xleftarrow{\text{Random}}$ Key Space
2) OBU: $Pkey_R^{+1} \longleftarrow H(ID_R)$
3) OBU: $\sigma \longleftarrow Encrypt_{Pkey_R^{+1}}(Cert, K_S^{+1})$
4) OBU $\longrightarrow$ RSU: $Sign_{Pkey_S^{-1}}(K_S^{+1}, \sigma)$
5) RSU: B $\longleftarrow$ Verify $(Decrypt_{Pkey_R^{-1}}(\sigma), K_S^{+1}, LR)$
6) RSU: If B=0 then exit
7) RSU: $ID_S \longleftarrow$ Random number
8) RSU: Cert $\longleftarrow Sign_{Pkey_R^{-1}}(K_S^{+1}||expiration||ID_S)$
9) RSU: Add $(K_S^{+1}, \sigma)$ to history table
10) RSU: Wait less than $\delta$ seconds
11) RSU $\longrightarrow$ OBU: Cert

Fig. 2. Certificate receiving algorithm.

is required for all messages (every persons must not access messages text). In the proposed protocol electronic signature is used for confidentiality. As a result, each receiver can identify transmitter of its received messages and transmitter can not deny that transmitting.

The proposed protocol is divided in two parts that use proper mechanisms due to theirs requirements and circumstances. Id-based electronic signature mechanisms is used in vehicle to infrastructure connection to satisfy vehicle tracking capability by reliable center and denial of service impossibility. A mechanism is required for legitimate vehicles identification and messages verification in vehicle to vehicle (V2V) connection due to vehicles limited sources and network economical requirements. So, a simple electronic signature protocol based on pseudonym is used for privacy protection in proposed protocol.

Major problems of previous protocols proposed for authentication and privacy protection in VANETs are high calculation and processing load that are forced to vehicles processors, consequently decreasing system efficiency and throughput severely. Solution of these problems is assignment of heavy calculation to network infrastructure as its possible or splitting calculations over the network. In this regard, a cooperative protocol is proposed to verifies each message signature that splits calculations between vehicles. Also, revoked certificates checking mechanism is done by roadside units. The proposed protocol include five subsequent steps.

*A. Network Set-up*

In the first step, each vehicle goes to a reliable center in person. The center issues a pair of public and private keys and a certificate for that vehicle after equipment technical checking. Also, it stores the keys and signed certificate on the vehicle and saves the signed certificate and vehicle identity in its database for future follow-up. Vehicle can use these certificate and keys to communicate with roadside units for a long time (e.g. a year).

Reliable center build a public and private key for each roadside unit depend on its zone. The roadside unit uses the private key for a period to communicate with vehicles. These keys are periodically updated after each technical checkpoint to prevent attacker manipulation.

*B. Temporary Certificate Recieving*

Every vehicle, in each zone, receives a temporary certificate from roadside unit to communicate with other vehicles that are in that zone. In this regard, vehicle should firstly authenticates itself to roadside unit. The roadside unit issues a temporary certificate after it makes sure the vehicle is legal and is not in the revoked list. When the vehicle certificate expired or it enter another roadside unit territory, it re-does this procedure to receive a new certificate.

As Fig. 2 shows vehicle S (OBU), in order to receive or update its temporary certificate, firstly build a pair of public and private keys $(K_S^{-1}, K_S^{+1})$ using electronic signature key building algorithm and signs the built public key using its private key. Then it uses a hash function and connects the roadside unit zone to achieves roadside unit public key and sends encrypted message to the relevant roadside unit (R). Roadside unit uses its private key to decrypt certificate request message received from vehicle S. Then it uses vehicle certificate to confirm it is not in the revoking list. Finally, roadside unit issues a temporary certificate by its private key. The certificate includes vehicle public key, certificate validity period, and an Id produced randomly by roadside unit. Roadside unit saves the built certificate with vehicle certificate on its database. Then it wait for at-most $\delta$ second to prevent detection of any relation between new and old certificates before it sent the certificate to vehicle S. Henceforth, vehicle S can communicate anonymously with other vehicles using this certificate.

## C. Message Signing and Sending

Each vehicle uses its private key ($K_S^{-1}$) to sign its messages. Then, it send the signed message with its temporary certificate and Id over the network.

## D. Signature Approval and Message Receiving

In previous proposed protocols for vehicle ad-hoc network, the vehicle checks each received message signature. Considering each vehicle sent a message in every 100-300 ms, and there were almost 100 vehicles at each vehicle communication coverage, each vehicle should check 1000 messages signature in just one second. So, the time for confirmation of each message should be less than 1 ms that is a very short time for electronic signature approval which cause receiver buffer overflow, and increase of messages loss rate. In this protocols, each message is checked by all vehicles receive it that strongly decrease the efficiency and throughput and cause vehicles processors be busy to do duplicated processes.

In the proposed protocol, each vehicle checks some of messages and informs other, in case of message signature was not verified. This solution reduces vehicles processing and computational load. So, strong processors are not required for vehicles result in system cost reduction. Also, high message loss rate due to electronic signature algorithms processing is reduced cause efficiency increasing.

The most important part of above solution is vehicles cooperation in order to achieve an acceptable rate of efficiency and throughput. In this protocol, each vehicle keeps a list of its neighbors Ids and up-date it with each message reception from the list member. Then it singes its neighbors list with its private key and send signed list with its temporary certificate over the network every $\theta$ seconds. Other vehicles use this list to identify their mutual neighbors with the sender, and check that are they the verifier of its messages or not? if vehicle A be the vehicle B verifier, since then it will verify vehicle B messages else it will wait for another vehicle to verify B. Consider $\beta$ as vehicle A & B Id difference, $\alpha_1$ as the highest and $\alpha_k$ as the shortest difference between vehicle B Id and its mutual neighbors with A, and $\alpha_p$ as the pth highest difference. In the proposed protocol, vehicle A verify B when, $\beta$ be less or equal $\alpha_p$.

Vehicle A receives message M from vehicle B and uses its Id that is in the M to check its neighbors list. If vehicle B Id wasn't in the list or vehicle A be the B verifier, checks vehicle V certificate by its public key. If the signature was verified, vehicle A update its list and delivers the message to application layer. On the other side, if the signature wasn't verified by vehicle B public key, vehicle A send a message over the network to informs others. In case vehicle A was not verifier of B, it waits for $\Delta t$ ms. If it receives a message in relation to message M disapproval, it will forward it to its neighbors and recheck message M signature for more assurance. If it didn't receive a message after $\Delta t$ ms telling message M disapproval, it considers M as a verified message and delivers it to application layer.

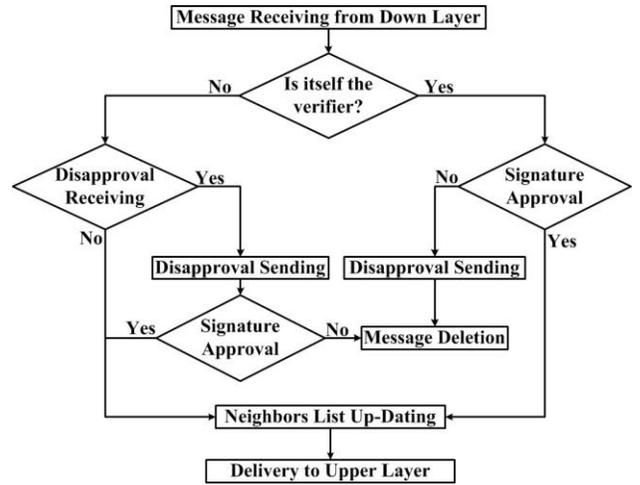

Fig. 3. Signature approval flowchart.

## E. Tracking and Certificate Revoking

Reliable center can extract the main certificate of vehicle off the roadside unit database using vehicle temporary certificate that is in its message. Then, the center can find malicious or attacker vehicle identity in its database. Finally, it add the target vehicle certificate to revoked certificates list and update it on all roadside units to prevent malicious vehicle from continuing its operation.

## III. ANALYSIS OF PROPOSED PROTOCOL

As it was said before, in previous protocols for VANETs each vehicle checks all received messages signature. It means each message is checked for n (number of vehicles received the message) times. While, in proposed protocol each message is checked at least p (not all vehicles received the message) times and at most 5×p times. The more p is decreased, less duplicated precesses are done by system and efficiency is increased. But, whatever p becomes lower, the probability of malicious abuses grows. So, there must be a trade off between efficiency and security.

The best case of choosing verifiers is that there be verifiers in both directions of vehicle. If p(B) be considered as the probability of choosing one vehicle in front and one vehicle in the back of vehicle to verify and N be the number of vehicles that should verify message M, for N=15 p(B) is equal to 0.99998 as shown in Fig. 4 that can be considered as 1 and is acceptable. So, the best value for p is 5 that each message is on average checked by 15 vehicles.

Another challenge in proposed protocol is determination of $\Delta t$. whatever this time gets longer, network end-to-end delay increases. On the other hand if this time wasn't enough, it's possible that an invalid message considered as valid. So, $\Delta t$ must bo longer than sum of times of one message confirmation and transmission delay. In the worth case, it is equal to message end-to-end delay when, the vehicle is checked all messages itself. Based on done simulations, this time average is equal to 15 ms when there are 25 vehicles around the vehicle. As a precaution Δt is considered 30 ms to assure all messages will be checked.

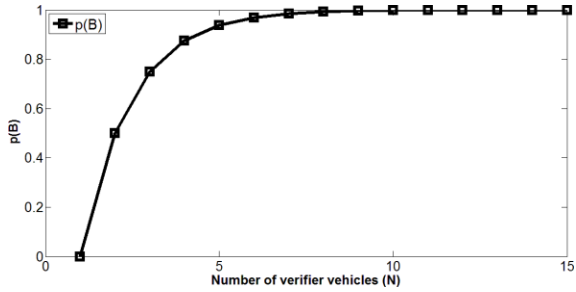

Fig. 4. Probability of choosing verifier vehicles in both sides of a vehicle.

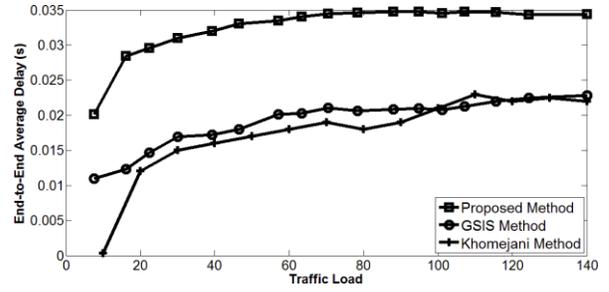

Fig. 6. Variation off latency respect to traffic load obtained from simulation.

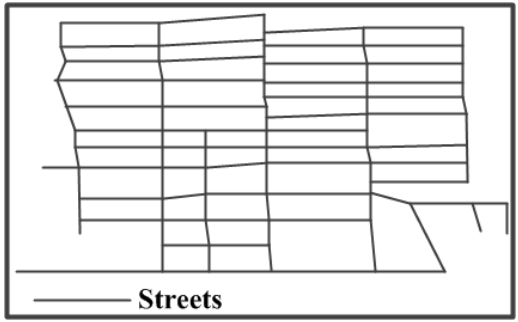

Fig. 5. Schematic of simulation zone.

Temporary certificate (key) can have several consequences on the proposed protocol. Shorter lifetime means better privacy protection and more on-time revoking. But, fast certificate updating is followed by higher connection and commuting costs. Also, some application programs of VANETs require messages relationship detection in a specified time, So the certificate lifetime must not be less than this time. Beside these, it should be noted that certificate lifetime is combination of time-based expiration and zone-based expiration, as a result lifetime selection should be proper to zone dimensions. Considering all above factors, 10 to 20 minutes is recommended for temporary certificate lifetime.

## IV. SIMULATION AND RESULT

A 3×3 $km^2$ urban area is considered as simulation zone, shown in Fig. 5. In simulation, Vehicles coverage is 300 meter that every 300 ms a message containing velocity, direction, acceleration, and location of vehicle is broadcasting.

Vehicle density is one the major factors have effect on system efficiency. Whatever vehicle density gets higher in a vehicle coverage, more vehicles receives transmitted message and also, the vehicle receives more messages. So, messages wait on the queue longer and end-to-end delay increases. Fig. 6 shows the effect of traffic load on end-to-end average delay for proposed protocol, and GSIS and Khomejani protocols.

As Fig. 6 shows, end-to-end delay for proposed protocol is almost fix despite the GSIS and Khomejani protocols that their delay increase with traffic load. The proposed protocol delay is longer than other two protocols caused by vehicle waiting for message confirmation by neighbors. This delay is predictable and is equal to summation of signing, message broadcasting, and waiting (Δt) times. It should be noted that in fact the simulated delay time (almost 35 ms) is very shorter than the acceptable delay value (100 ms) defined in IEEE1609 standard. So, from the standpoint of delay, the proposed protocol will mack no problem for application programs.

Traffic load effect on messages-loss average rate for the proposed protocol, and the GSIS and Khomejani ones are shown in Fig. 7. It is seen that in GSIS and Khomejani protocols message loss rate increases with traffic load, but the loss rate for proposed protocol is almost zero. This is the result of vehicles cooperation for messages signature checking.

Number of checked signature in proportion with received messages in the proposed protocol can be seen in Fig. 8. The more this value approaches to 1, it means vehicles cooperation decreases and more duplicated works is done in the network. In the proposed protocol, whatever traffic load gets higher, cooperation increase and each vehicle checks less messages.

Also, number of messages checked by one vehicle in 30 ms period is illustrated in Fig. 9. Vehicle received messages increase with traffic load. In the GSIS protocol, the processor can only check 43 messages in every 300 ms, so in every period (300 ms) at most 43 messages is delivered to upper layer. But, vehicles cooperation in proposed protocol causes the processor checks fewer messages and loaded with less processing load. So, the proposed protocol has a better efficiency than the GSIS protocol.

TABLE III
SIMULATION SETTING

| Abbreviations | Explanation |
| --- | --- |
| Topology dimensions | 3000×3000 m |
| Simulation time | 100 s |
| Vehicle coverage | 300 m |
| Network bandwidth | 6 Mb/s |
| Access layer channel protocol | IEEE 802.11 P |
| Vehicles antenna | Omni-antenna |
| Vehicles speed | 30-75 Km/h |

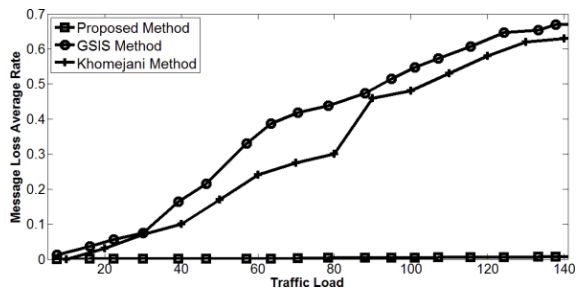

Fig. 7. Estimated traffic message loss ratio versus different traffic loads.

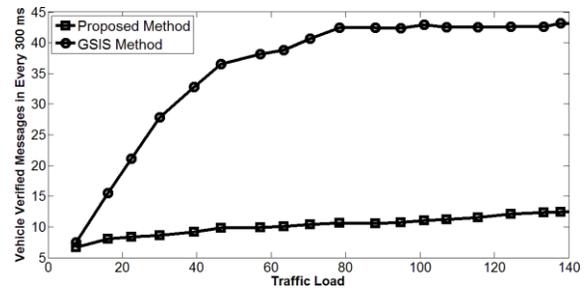

Fig. 9. Messages verified in 300 ms by every vehicle on proposed and GSIS protocols.

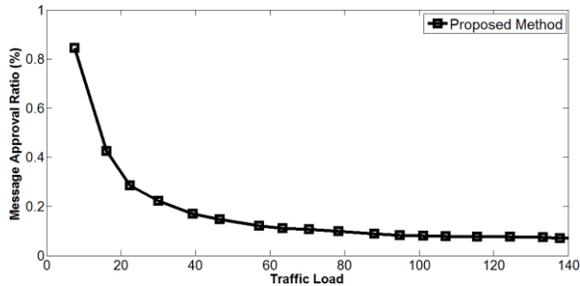

Fig. 8. Message approval ratio obtained from simulation against traffic load.

## V. CONCLUSION

In this paper, a new protocol is proposed for user authentication and privacy protection. In this protocol, vehicles co- operation is used for messages signature checking to decrease messages loss rate. The simulations results show significant improvement on message loss rate parameter with comparison to the GSIS and the Khomejani protocols. On the other hand, end-to-end delay of proposed protocol is longer than the other ones, but it's still much less than the maximum acceptable value defined by standards and could be overlooked.